\documentclass{article}
\usepackage{amsfonts}
\usepackage{amsmath}

\begin{document}

\title{Ward-like Operator in the comma theory}
\author{A Abdurrahman\thanks{%
Ababdu@ship.edu} \\
Department of Physics\\
Shippensburg University of Pennsylvania\\
1871 Old Main Drive\\
Shippensburg, PA 17257\\
USA}
\maketitle

\begin{abstract}
The construction of the Ward-like operator of the interacting string field
theory presented by Witten, in terms of the half-string (comma) oscillator
modes, is completed.
\end{abstract}

\section{Introduction}

\bigskip Witten's cubic bosonic open string field theory was formulated in 
\cite{1} and made more precise in terms of full string oscillators in \cite%
{Gross-Jevicki-I-II,T MORRIS}. In \cite{1}, Witten suggested that
the full string ($FS$) may be split into two halves leading to the\
half-string (comma) approach formulated in \cite{HMCHAN1}. In \cite{Bordes}
a precise formulation in terms of half-string oscillators was given. Further
development of the comma theory was carried out in \cite{AbdulI-II-III}. In
2001, the work of references \cite%
{Sen-Zwiebach,Rastelli-Sen-Zwiebach,Gross-Taylor-I,Gross-Taylor-II}\
generated much interest in the comma formulation of Witten's theory of
interacting open bosonic strings. Subsequently, interest in the comma theory
faded a way and many of the issues in the comma approach raised in \cite%
{HSBOSEFERMI,AbdulI-II-III} remain unsolved. In \cite{Hlousek-Jevicki},
Ward-like ($\mathcal{WL}$) identities were established and used as part of
the proof of the equivalence between the bosonic and fermionic ghost
realizations of the gauge invariance in Witten's theory. Although $K$
invariance was established in \cite{HSBOSEFERMI}, the $\mathcal{WL}$
identities were not discussed in the case of the comma theory. Just as in
the case of the full string, the proof of equivalence in the comma theory
requires that similar $\mathcal{WL}$ identities hold in the comma theory.
The importance of the proof resides in the fact that the physical content of
the theory depends on $\mathcal{WL}$ identities. One of the reasons that
this problem received little attention has to do with the difficulty of
constructing the $\mathcal{WL}$ operator in the comma approach as we shall
see shortly. Once we construct the $\mathcal{WL}$ operator, we can complete
the proof of equivalence of the comma three-point vertex (coordinate and
ghost), and establish the BRST invariance of the full three-point vertex in
the bosonized ghost\cite{AbduWardId}, a problem that was briefly discussed
in \cite{HSBOSEFERMI}.

\section{\protect\bigskip Comma formalism}

Here we shall review the comma formalism needed to describe the $\mathcal{WL}
$ operator in terns of the half-string oscillator modes. The comma formalism
was developed in \cite{Bordes}. Here we briefly consider the bosonic ghost
field $\phi\left( \sigma\right) $; the matter fields $x^{\mu\text{ }}\left(
\sigma\right) $, $\mu=0,1,2,...,25$, have similar structure and have been
considered in \cite{Bordes}.

\subsection{Full string ghost field}

\bigskip The bosonized ghost field $\phi\left( \sigma\right) $ for the full
string has the mode expansion%
\begin{equation}
\phi\left( \sigma\right) =\phi_{0}+\sqrt{2}\sum_{n=1}^{\infty}\phi_{n}\cos
n\sigma  \label{eqn2}
\end{equation}
where \ $0\leq\sigma\leq\pi$. In (\ref{eqn2}) the modes $\phi_{n}$ are
related to creation and annihilation operators through 
\begin{align}
\phi_{n} & =\frac{i}{\sqrt{2n}}\left( a_{n}^{\phi}-a_{-n}^{\phi}\right) \ \
\ \ \ \ \ \ \ \ \  \\
p_{n}^{\phi} & =-i\frac{\partial}{\partial\phi_{n}}=\sqrt{\frac{n}{2}}\left(
a_{n}^{\phi}+a_{-n}^{\phi}\right)
\end{align}
for $n>0$ (with $a_{-n}^{\phi}\equiv a_{n}^{\phi}$ ) and 
\begin{align}
\phi_{0} & =\frac{i}{2}\left( a_{0}^{\phi}-a_{0}^{\phi\dagger}\right) \ \ \
\ \ \ \ \ \  \\
\ \ p_{0}^{\phi} & =-i\frac{\partial}{\partial\phi_{0}}=\left( a_{0}^{\phi
}+a_{0}^{\phi\dagger}\right)
\end{align}
for $n=0$.

\subsection{\protect\bigskip Comma ghost field}

Here we review the transformation needed to express the full string ghost
field in terms of ghost fields living on the left and right halves of the
string. Following ref. \cite{Bordes}, we break up the full string ghost
coordinate $\phi\left( \sigma\right) $ satisfying Neumann boundary
conditions at the end points of the string ($\sigma=0,\pi)$ into its left
and right halves, according to%
\begin{equation}
\varphi^{\left( L\right) }\left( \sigma\right) =\phi(\sigma)-\phi (\frac{\pi%
}{2}),\text{ \ \ \ }0\leq\sigma\leq\frac{\pi}{2}\text{\ \ }
\label{defofhscooghost-L}
\end{equation}%
\begin{equation}
\varphi^{\left( R\right) }\left( \sigma\right) =\phi(\pi-\sigma )-\phi(\frac{%
\pi}{2}),\text{ \ }0\leq\sigma\leq\frac{\pi}{2}  \label{defofhscooghost-R}
\end{equation}
where now the comma ghost fields $\varphi^{\left( L\right) }\left(
\sigma\right) $, $\varphi^{\left( R\right) }\left( \sigma\right) $ obeys
Neumann boundary condition at $\sigma=0$ and Dirichlet boundary condition at 
$\sigma=\pi/2$. The mode expansion of the left and right pieces have been
considered in \cite{Bordes}%
\begin{equation}
\varphi^{\left( r\right) }\left( \sigma\right) =\sqrt{2}\sum_{n=1}^{\infty}%
\varphi_{n}^{\left( r\right) }\cos\left( 2n-1\right) \sigma\text{ }
\label{eq3}
\end{equation}
where $r=1,2=L,R$. The modes $\varphi_{n}^{\left( r\right) }$ in (\ref{eq3})
are related to the comma creation and annihilation operators through%
\begin{equation}
\varphi_{n}^{\left( r\right) }=\frac{i}{\sqrt{2n-1}}\left(
b_{n}^{\varphi\left( r\right) }-b_{-n}^{\varphi\left( r\right) }\right) 
\text{\ \ \ \ \ \ \ \ \ }
\end{equation}%
\begin{equation}
\text{\ \ \ \ \ \ \ }p_{n}^{\varphi\left( r\right) }=-i\frac{\partial }{%
\partial\varphi_{n}^{\left( r\right) }}=\frac{\sqrt{2n-1}}{2}\left(
b_{n}^{\varphi\left( r\right) }+b_{-n}^{\varphi\left( r\right) }\right) 
\text{ }
\end{equation}
for $n>0$ (with $b_{-n}^{\left( r\right) }\equiv b_{n}^{\left( r\right)
\dagger}$). The degrees corresponding to the midpoint ghost coordinate and
and its conjugate momenta are given by%
\begin{equation*}
\varphi_{M}\equiv\phi\left( \pi/2\right) \ \ \ \ \ \ \ \ 
\end{equation*}%
\begin{equation}
=\phi_{0}+\frac{\sqrt{2}}{\pi}\sum_{n=1}^{\infty}\frac{\left( -1\right) ^{n}%
}{2n-1}\left( \varphi_{n}^{\left( 1\right) }\ +\varphi_{n}^{\left( 2\right)
}\right) \ \ \ \ \ \ \ \ 
\end{equation}
and 
\begin{equation}
\ \ \wp_{M}^{\varphi}=-i\frac{\partial}{\partial\phi\left( \pi/2\right) }%
=p_{0}^{\phi}\text{,}
\end{equation}
respectively.

The operators $b^{\prime}s$ are related to the usual comma operators $%
\beta^{\prime}s$ (corresponds in the full string to $\alpha^{\prime}s$)
through 
\begin{equation}
\text{ \ }b_{n}^{\left( r\right) }=\frac{\beta_{2n-1}^{\varphi\left(
r\right) }}{\sqrt{2n-1}}\text{,\ \ \ \ \ \ \ \ }b_{-n}^{s}=\frac{\beta
_{2n-1}^{\varphi\left( r\right) \dagger}}{\sqrt{2n-1}}\text{\ \ \ }
\end{equation}
The full string ghost operators $\alpha_{n}^{\phi}$ are related to the comma
ghost operators $\beta_{n}^{\varphi}$ through \cite{AbdulI-II-III}.%
\begin{equation}
\alpha_{2n}^{\phi}=\displaystyle\sum_{m=1}^{\infty}M_{2n\text{ }%
2m-1}\beta_{2m-1}^{\varphi\left( +\right) }+\left( -1\right)
^{n}\wp_{M}^{\varphi}\text{,\ \ \ \ \ }
\end{equation}%
\begin{equation}
\text{ \ \ \ \ \ }\alpha_{2n-1}^{\phi}=\beta_{2n-1}^{\varphi\left( +\right) }%
\text{, \ \ \ }
\end{equation}
for $n>0$, and 
\begin{equation}
\alpha_{0}^{\phi}=p_{0}^{\phi}=\wp_{M}^{\varphi}  \label{eqnchangeofbhf}
\end{equation}
for $n=0$. The operators \ $\beta_{n}^{\left( \pm\right) }$ are defined by 
\begin{equation}
\text{ \ }\beta_{n}^{\left( \pm\right) }=\frac{1}{\sqrt{2}}\left( \beta
_{n}^{\left( 1\right) }\pm\beta_{n}^{\left( 2\right) }\right) \text{\ }
\end{equation}
and the change of representation matrix, $M$, has the form 
\begin{equation}
\text{ \ }M_{nm}\text{\ \ }=\frac{n}{m}\text{ }\frac{\left( -1\right)
^{\left( n+m+1\right) /2}}{n-m}\text{\ }
\end{equation}

\subsection{The three vertex in the comma theory}

In the comma approach to string theory, the elements of the theory are
defined by $\delta-function$ type overlaps%
\begin{equation}
V_{3}^{\left( \chi+\varphi\right) HS}=\exp^{i\sum_{j=1}^{3}Q_{j}^{\left(
\phi\right) }\phi\left( \pi/2\right) }V_{3,0}^{\left( \varphi\right)
HS}V_{3,0}^{\left( \chi\right) HS}
\end{equation}
where%
\begin{equation}
V_{3,0}^{\left( \varphi\right) HS}=\displaystyle\prod_{j=1}^{3}\displaystyle%
\prod \limits_{\sigma=0}^{\pi/2}\delta\left[ \varphi_{j}^{\left( L\right)
}\left( \sigma\right) -\varphi_{j-1}^{\left( R\right) }\left( \sigma\right) %
\right]
\end{equation}
and $V_{3,0}^{\left( \chi\right) HS}$ refers to the matter part of the
vertex \cite{AbdulI-II-III}. It is to be understood that $j-1=0\equiv3$. The
factor $Q_{j}^{\left( \phi\right) }$ is the ghost number insertion at the
mid-point which is needed for the $BRST$ invariance of the theory \cite%
{Gross-Jevicki-I-II,Bordes} and in this case $Q_{1}^{\left( \phi\right)
}=Q_{2}^{\left( \phi\right) }=Q_{3}^{\left( \phi\right) }=1/2$. As we have
seen before \cite{Gross-Jevicki-I-II,Bordes}, in the Hilbert space of the
theory, the $\delta-functions$ overlaps translate into operator overlap
equations which determine the precise form of the vertex. The ghost part of
the comma vertex has the same structure as the coordinate one apart from the
mid-point insertions 
\begin{equation}
\left\vert V_{3}\right\rangle _{HS}^{\chi+\varphi}=e^{\frac{3}{2}%
i\phi_{M}}V_{3,0}^{\left( \varphi\right) HS}V_{3,0}^{\left( \chi\right)
HS}\left\vert 0,N_{g}=\frac{3}{2}\right\rangle _{HS}^{\chi+\varphi }
\label{eqnGhost3vertexHS}
\end{equation}
where 
\begin{equation}
\varphi_{M}\equiv\phi_{1}\left( \frac{\pi}{2}\right) =\phi_{2}\left( \frac{%
\pi}{2}\right) =\phi_{3}\left( \frac{\pi}{2}\right)
\end{equation}
and%
\begin{equation*}
\left\vert 0,N_{g}=\frac{3}{2}\right\rangle _{HS}^{\chi+\varphi}=\left\vert
0\right\rangle _{L}^{\chi}\left\vert 0\right\rangle _{R}^{\chi}\times
\end{equation*}%
\begin{equation}
\left\vert 0,N_{g}=\frac{3}{2}\right\rangle _{L}^{\varphi}\left\vert 0,N_{g}=%
\frac{3}{2}\right\rangle _{R}^{\text{ }\varphi}
\end{equation}
In the above expression, $\left\vert 0\right\rangle _{123}=\left\vert
0\right\rangle _{1}\left\vert 0\right\rangle _{2}\left\vert 0\right\rangle
_{3}$ and $V_{3,0}^{HS}$ is the exponential of a quadratic form in the
creation operators 
\begin{equation}
V_{3,0}^{HS}=\exp\left[ \frac{1}{2}\frac{\beta_{n}^{\dag r\text{ }i}}{\sqrt{n%
}}H_{nm}^{ij;ss^{\prime}}\frac{\beta_{m}^{\dag s\text{ }j}}{\sqrt{m}}\right]
\end{equation}
The matrix elements $H_{nm}^{ij;ss^{\prime}}$ have the simple form%
\begin{equation}
H_{nm}^{rs;\text{ }ij}=2\delta_{nm}\delta^{rR}\delta^{SL}\delta^{i\text{ }%
j-1}
\end{equation}
with $i-1=0\equiv3$ and the half string operators $\beta^{\prime}s$ satisfy
the usual commutation relations%
\begin{equation}
\text{ }\left[ \beta_{n}^{r\text{ }i},\beta_{m}^{\dag s\text{ }j}\right]
=n\delta^{rs}\delta^{ij}\delta_{n,m}\text{, \ \ }n,m\text{ odd}
\end{equation}
It is the simplicity of the coupling matrix, $H$, that makes the comma
theory appealing (Compare $H$ to $N$ in ref. \cite{Gross-Jevicki-I-II}).

\section{$\mathcal{WL}$ operator in the FS basis}

\bigskip We recall that in ref. \cite{Gross-Jevicki-I-II,T MORRIS} the
bosonic and fermionic realizations of the full string ghost fields have been
used to express the ghost part of the string field but no attention was paid
to their equivalence. In ref. \cite{Gross-Jevicki-I-II}, the authors showed
that both ways give rise to a gauge invariant theory for Witten's open
bosonic string. However, their equivalence was never established there and
the proof of equivalence of the two formulations of the ghost part was
subsequently established in ref. \cite{Hlousek-Jevicki}. In ref. \cite%
{Hlousek-Jevicki} it was established that operator vertices obey the same
equations and are identical. The key to the proof lies in the various
identities satisfied by the Neumann coefficients in the definition of the
three point vertex. In the proof of equivalence the authors needed to
establish $K$-invariance and $\mathcal{WL}$ identities ($L$-equations) which
are necessary but not sufficient for gauge invariance which hods only in $%
D=26$. They also proved $BRST$-invariance and showed that $BRST$-invariance
implies both $K$-invariance and the $WL$ identities. In \cite{HSBOSEFERMI},
the $K$-invariance of the comma theory was established and the $BRST$%
-invariance was briefly discussed, but the $\mathcal{WL}$ identities were
not addressed. Before discussing the $\mathcal{WL}$ identities and $BRST$%
-invariance in the comma theory one needs, first, to construct the $\mathcal{%
WL}$ operator in the comma theory. Before we construct the $\mathcal{WL}$
operator in the comma theory, let us briefly review the $\mathcal{WL}$
identities in the full string oscillator formalism of the open bosonic
string.

In the bosonic representation of the ghost, the $\mathcal{WL}$ identities
for the Witten vertex matter plus ghost \cite{Hlousek-Jevicki} are given by%
\begin{equation}
W_{m}^{x+\phi,i}||V_{3}^{x+\phi}>=0\text{, \ \ }m=1,2,...\text{; }i=1,2,3
\label{eqnWardIdBose}
\end{equation}
where, $W_{m}^{x+\phi}$, a Ward-like operator, is defined by 
\begin{equation}
W_{m}^{x+\phi,i}=L_{m}^{x+\phi,i}+\sum_{j=1}^{3}\sum_{n=0}^{\infty }m%
\widetilde{N}^{ij}L_{-n}^{x+\phi,j}
\end{equation}
The full Virasoro generators, $L_{m}^{x+\phi}$ consist of $L_{m}^{x}$ +$%
L_{m}^{\phi}$ , where the ghost Virasoro generators are%
\begin{equation*}
L_{m}^{\phi}=\displaystyle\sum
\limits_{n=1}^{\infty}\alpha_{n}^{\phi\dag}\alpha_{n+m}^{\phi}+\frac{1}{2}%
\displaystyle\sum
\limits_{n=1}^{m-1}\alpha_{m-n}^{\phi}\alpha_{n}^{\phi}+p_{0}^{\phi}%
\alpha_{m}^{\phi}
\end{equation*}%
\begin{equation}
-\frac{3}{2}m\alpha_{m}^{\phi}\text{, \ \ }m>0  \label{eq3vgfoeam}
\end{equation}
and%
\begin{equation}
L_{0}^{\phi}=\displaystyle\sum
\limits_{n=1}^{\infty}\alpha_{n}^{\phi\dag}\alpha_{n}^{\phi}+\frac{1}{2}%
\left( p_{0}^{\phi}\right) ^{2}-\frac{1}{8}
\end{equation}
Notice that $L_{m}^{\phi}$ differs from the coordinate piece $L_{m}^{x}$
only in the extra term, linear in $\alpha_{m}^{\phi}$, arising because of
the extra term in the energy-momentum tensor of $\phi(\sigma)$ 
\begin{equation}
T_{\pm}=\frac{1}{2\pi}\left[ \left( \partial_{\pm}\phi\right) ^{2}-\frac {3}{%
2}\partial_{\pm}^{2}\phi\right]  \label{eqenergymomtensorbose}
\end{equation}
It is this extra term that gives rise to $\mathcal{WL}$ identities in 26
dimensions. With this extra term the Virasoro generators for the ghost
satisfy 
\begin{equation}
\left[ L_{n}^{\phi},L_{m}^{\phi}\right] =\left( n-m\right) L_{n+m}^{\phi
}-\left( n-m\right) \delta_{n+m}
\end{equation}
The constant term in the commutator $\left[ L_{n}^{\phi},L_{m}^{\phi}\right] 
$ is missing from the commutator for the matter part $\left[
L_{n}^{x},L_{m}^{x}\right] $. Likewise the zero mode $L_{0}^{\phi}$ differs
from the matter part $L_{0}^{x}$ in the constant $-1/8$. This extra term is
needed since the vacuum ghost number has the value $p^{\phi}=1/2$ and
therefore $L_{0}^{\phi}=0$ as desired.

\subsection{\protect\bigskip$\mathcal{WL}$ identities in the comma basis}

\bigskip In the bosonization of the fermionic coordinates for the comma
string, using the standard procedure (see Ref. 1 in \cite{T MORRIS}), it is
not obvious that all ingredients of the comma theory employing the comma
bosonic fields $\varphi^{L}\left( \sigma\right) $ and $\varphi^{R}\left(
\sigma\right) $ are equivalent to those constructed using the original comma
fermi fields $c^{L}\left( \sigma\right) $, $c^{R}\left( \sigma\right) $, $%
b^{L}\left( \sigma\right) $ and $b^{R}\left( \sigma\right) $ appearing in 
\cite{Bordes}. It has been claimed in \cite{HSBOSEFERMI}, that the ghost
bosonic vertices in the comma operator formulation obey the same overlap
equations as the fermionic vertices and are identical. However a rigorous
proof was not presented in \cite{Bordes}. The authors failed to establish
that the comma ghost (plus matter) vertex in the bosonic realization of the
ghosts satisfy the same $\mathcal{WL}$ identities obeyed by the comma ghost
(plus matter) vertex in the fermionic realization of the ghosts. To complete
the proof of equivalence between the two realization of the comma ghost
vertex, we need to establish the $\mathcal{WL}$ identities utilizing the
bosonic representation of the comma ghost fields as well. But to do that one
first needs the explicit form of the $\mathcal{WL}$ operator in the comma
formalism. Finding the explicit form of the $\mathcal{WL}$ operator in the
comma formalism will be the subject of this paper. In a sequel to this paper
we will prove that $\mathcal{WL}$ identities do indeed hold in the comma
theory of the open bosonic string and use them to establish $K$-invariance 
\cite{AbduWardId}. However, as it is the case in the full string, a direct
proof of the $BRST$-invariance that follow from the $\mathcal{WL}$
identities is not possible due to the presence of the $\left( 1/2\right)
L^{\phi,i}$ term in the $BRST$ charge. In \cite{AbduWardId}, we will see
that the $\mathcal{WL}$ identities and the $K$-invariance are necessary but
not sufficient conditions for gauge invariance. On the other hand we will
see that $BRST$-invariance implies both the $\mathcal{WL}$ identities and
the $K$-invariance \cite{AbduWardId}.

To prove the equivalence between the two ghost realizations in the comma
theory we need to show that a similar identity to the one in (\ref%
{eqnWardIdBose}) holds in the comma basis; that is, we need to show that%
\begin{equation}
\left[ \mathcal{W}_{m}^{\chi+\varphi,i}\left( L\right) +\mathcal{W}%
_{m}^{\chi+\varphi,i}\left( R\right) \right] |\left( V_{3}\right)
_{HS}^{\chi+\varphi}>=0\text{\ }
\end{equation}
holds for $m\geq0$ and $i=1,2,3$. The operators $W\left( L\right) $ and $%
W\left( R\right) $ refer to the left and right parts of the $\mathcal{WL}$
operators, respectively, and are given by 
\begin{equation}
\mathcal{W}_{m}^{\chi\left( \varphi\right) ,i}\left( L\right) =\ell
_{m}^{\left( L\right) \chi\left( \varphi\right)
,i}+\sum_{j=1}^{3}\sum_{n=0}^{\infty}m\widetilde{N}_{mn}^{ij}\ell_{-n}^{%
\left( L\right) \chi\left( \varphi\right) ,j}
\end{equation}
and 
\begin{equation}
\mathcal{W}_{m}^{\chi\left( \varphi\right) ,i}\left( R\right) =\ell
_{m}^{\left( R\right) \chi\left( \varphi\right)
,i}+\sum_{j=1}^{3}\sum_{n=0}^{\infty}m\widetilde{N}_{mn}^{ij}\ell_{-n}^{%
\left( R\right) \chi\left( \varphi\right) ,j}\text{,}
\end{equation}
for $m\geq0$, respectively. The operators $\ell_{-n}^{\left( L\right) }$ and 
$\ell_{-n}^{\left( R\right) }$ are the comma Virasoro generators
corresponding to the left and right parts of the string. One way to obtain
the comma ghost Virasoro generators, we recall that the full string Virasoro
generators are defined as%
\begin{equation}
L_{m}=\frac{1}{4\pi}\int_{-\pi}^{\pi}d\sigma e^{im\sigma}\left. \left[ 
\overset{\cdot}{\phi}\left( \sigma,\tau\right) +\phi^{\prime}\left(
\sigma,\tau\right) \right] ^{2}\right\vert _{\tau=0}  \label{eqndefofviraop}
\end{equation}
To use the change of representation from the full string to the comma string
in (\ref{defofhscooghost-L}) and (\ref{defofhscooghost-R}), we need to
reduce the range of integration to $\sigma\in\left[ 0,\pi/2\right] $. Thus%
\begin{equation*}
L_{m}=\frac{1}{4\pi}\int_{0}^{\pi/2}d\sigma\left[ e^{im\sigma}\left( \overset%
{\cdot}{\phi}\left( \sigma\right) +\phi\left( \sigma\right) ^{\prime}\right)
^{2}\right.
\end{equation*}%
\begin{equation*}
\left. +e^{-im\sigma}\left( \overset{\cdot}{\phi}\left( \sigma\right)
-\phi^{\prime}\left( \sigma\right) \right) ^{2}\right]
\end{equation*}%
\begin{equation}
+\frac{\left( -1\right) ^{m}}{4\pi}\int_{0}^{\pi/2}d\sigma\left[
e^{im\sigma}\left( \overset{\cdot}{\phi}\left( \pi-\sigma\right) +\phi\left(
\pi-\sigma\right) ^{\prime}\right) ^{2}\right.  \notag
\end{equation}%
\begin{equation}
\left. +e^{-im\sigma}\left( \overset{\cdot}{\phi}\left( \pi-\sigma\right)
-\phi^{\prime}\left( \pi-\sigma\right) \right) ^{2}\right]
\label{eqndefofviraopLR}
\end{equation}
Consistency with the definition of the comma string in (\ref%
{defofhscooghost-L}) and (\ref{defofhscooghost-R}), requires that we
identify left comma Virasoro generators by 
\begin{equation*}
\ell_{m}^{\left( L\right) }=\frac{1}{4\pi}\int_{0}^{\pi/2}\left[
e^{im\sigma}\left( \overset{\cdot}{\phi}\left( \sigma\right) +\phi\left(
\sigma\right) ^{\prime}\right) ^{2}\right.
\end{equation*}%
\begin{equation}
\left. +e^{-im\sigma}\left( \overset{\cdot}{\phi}\left( \sigma\right)
-\phi^{\prime}\left( \sigma\right) \right) ^{2}\right] d\sigma
\label{eqndefofviraopLeft}
\end{equation}
and identify right comma Virasoro generators by 
\begin{equation*}
\ell_{m}^{\left( R\right) }=\frac{\left( -1\right) ^{m}}{4\pi}%
\int_{0}^{\pi/2}\left[ e^{im\sigma}\left( \overset{\cdot}{\phi}\left(
\pi-\sigma\right) +\phi\left( \pi-\sigma\right) ^{\prime}\right) ^{2}\right.
\end{equation*}%
\begin{equation}
\left. +e^{-im\sigma}\left( \overset{\cdot}{\phi}\left( \pi-\sigma\right)
-\phi^{\prime}\left( \pi-\sigma\right) \right) ^{2}\right] d\sigma
\label{eqndefofviraopRight}
\end{equation}
The full Virasoro generators do not factorize completely into left and right
parts. This is because of the midpoint in the definition of the comma ghost
coordinates in (\ref{defofhscooghost-L}) and (\ref{defofhscooghost-R}) since 
$\varphi^{\left( L\right) }\left( \pi/2\right) =\varphi^{\left( R\right)
}\left( \pi/2\right) =0$. One can avoid this problem by execluding the
midpoint $\phi\left( \pi/2\right) $ from the definition of the half-string
ghost coordinates in (\ref{defofhscooghost-L}) and (\ref{defofhscooghost-R}%
). If one excludes the midpoint from the definition of the half string ghost
coordinates the Virasoro generators will separate into left and right
Virasoro generators with the constraint $\lim_{\sigma\rightarrow\pi/2}%
\varphi^{\left( L\right) }\left( \sigma\right)
=\lim_{\sigma\rightarrow\pi/2}\varphi^{\left( R\right) }\left( \sigma\right) 
$. However, we need not worry about that here. Thus using the definitions in
(\ref{eqndefofviraopLeft}) and (\ref{eqndefofviraopRight}), equation (\ref%
{eqndefofviraopLR}) can be written in the form 
\begin{equation}
L_{m}^{i}=\ell_{m}^{\left( L\right) ,i}+\ell_{m}^{\left( R\right) ,i}
\end{equation}
Henceforth, when the Virasoro generators are expressed in terms of the full
string modes we shall write $L_{m}$ and when they are expressed in terms of
the comma modes we shall write $\ell_{m}=\ell_{m}^{\left( L\right)
}+\ell_{m}^{\left( R\right) }$.

The full string vacuum is related to the comma vacua through the relation%
\begin{equation}
\left\vert |0\right\rangle =\exp\left( -\frac{1}{2}\frac{\beta_{2n-1}^{%
\left( +\right) \dag}}{\sqrt{2n-1}}\Phi_{mn}\frac{\beta_{2m-1}^{\left(
+\right) \dag}}{\sqrt{2m-1}}\right) \left\vert 0\right\rangle _{L}\left\vert
0\right\rangle _{R}  \label{eqnfvactohalfvac}
\end{equation}
where $\left\vert 0\right\rangle _{L\left( R\right) }$ correspond to the
comma left (right) vacuum. The operators $\beta^{\left( +\right) \prime}s$
are defined by%
\begin{equation}
\beta_{2n-1}^{\left( +\right) }=\frac{1}{\sqrt{2}}\left( \beta
_{2n-1}^{\left( L\right) }+\beta_{2n-1}^{\left( R\right) }\right)
\end{equation}
and the matrix $\Phi$ has been defined in reference \cite{Bordes} 
\begin{equation}
\Phi_{mn}=\frac{\sqrt{2m-1}\sqrt{2n-1}}{2\left( m+n-1\right) }\left( 
\begin{array}{c}
-\frac{1}{2} \\ 
m-1%
\end{array}
\right) \left( 
\begin{array}{c}
-\frac{1}{2} \\ 
n-1%
\end{array}
\right)
\end{equation}

\subsection{The comma WL operator}

To construct the comma $\mathcal{WL}$ operator, we first need to derive the
Virasoro generators in the comma formalism of string theory. The matter part
has been constructed in \cite{Bordes}; hence here we shall focus on the
ghost part of the Virasoro generators.

When quantizing a theory one has to deal with normal ordering ambiguities.
In the full string theory the vacuum state $||0>$ is related to the left and
right vacuum states $|0>_{L}$and $|0>_{R}$ through (\ref{eqnfvactohalfvac}).
In the full string formalism \cite{Bordes} one is interested in normal
ordering with respect to the full string vacuum (\ref{eqnfvactohalfvac}) but
in the comma theory formulated in \cite{HMCHAN1}, the relevant normal
ordering would be with respect to the comma left and right vacuum states $%
|0>_{L}$and $|0>_{R}$. However, the two criteria are linked through the
change of representation relationships.

When ordering operators, we need to regularize the divergent quantities
resulting from normal ordering. One way of doing that is based on the so
called "heat kernel regularization," in which one replaces the operators $%
\alpha_{n}^{\prime}s$ and $\beta_{n}^{\prime}s$ by

\begin{equation}
\alpha_{n}^{\epsilon}=e^{-\left[ \frac{n+1}{2}\right] \epsilon}\alpha _{n}%
\text{, \ \ \ \ \ }\epsilon>0
\end{equation}%
\begin{equation}
\beta_{n}^{\epsilon}=e_{n}^{-\left[ \frac{n+1}{2}\right] \epsilon}\beta _{n}%
\text{, \ \ \ \ \ }\epsilon>0
\end{equation}
so that a given operator in any of the representations can be obtained by a
limiting procedure for infinitesimally small $\epsilon$. Thus for a given
operator $O$ we write%
\begin{equation*}
O\left( \alpha_{n}\right) =\lim_{\epsilon\rightarrow0}O\left( \alpha
_{n}^{\epsilon}\right)
\end{equation*}%
\begin{equation}
=\lim_{\epsilon\longrightarrow0}O\left( e^{-\left[ \frac{n+1}{2}\right]
\epsilon}\alpha_{n}\right)
\end{equation}
for the full string oscillators and 
\begin{equation*}
O\left( \beta_{n}^{\left( r\right) }\right) =\lim_{\epsilon\rightarrow
0}O\left( \beta_{n}^{\left( r\right) \epsilon}\right)
\end{equation*}%
\begin{equation}
=\lim_{\epsilon\longrightarrow0}O\left( e_{n}^{-\left[ \frac{n+1}{2}\right]
\epsilon}\beta_{n}^{\left( r\right) }\right)
\end{equation}
for the comma oscillators ($r=L,R$). Using this prescription it is now
possible to find the relationship between the two normal orderings. For
example,%
\begin{equation*}
::\alpha_{n}^{\epsilon}\alpha_{m}^{\epsilon}::=:\alpha_{n}^{\epsilon}%
\alpha_{m}^{\epsilon}:+\frac{1+\left( -1\right) ^{n}}{2}
\end{equation*}%
\begin{equation}
\times\frac{1+\left( -1\right) ^{m}}{2}\left[ K_{nm}-\delta_{N+m,0}\theta%
\left( m\right) \right]
\end{equation}
where in the above identity $::$ $::$ and $::$ indicate normal ordering with
respect to the full string and comma vacuum states, respectively.

To first order in $\epsilon$, the quantity $K_{nm}$ reads \cite{Bordes}, 
\begin{equation*}
K_{nm}=-\frac{1}{2}\frac{nm}{n+m}\left\{ \frac{1}{n}\left[ \psi\left( \frac{1%
}{2}\right) -\psi\left( \frac{1+\left\vert n\right\vert }{2}\right) \right]
\right.
\end{equation*}%
\begin{equation}
+\frac{1}{m}\left[ \psi\left( \frac{1}{2}\right) -\psi\left( \frac{%
1+\left\vert m\right\vert }{2}\right) \right]  \notag
\end{equation}%
\begin{equation}
\left. \epsilon\left[ \psi\left( \frac{1+\left\vert n\right\vert }{2}\right)
-\psi\left( \frac{1+\left\vert m\right\vert }{2}\right) \right] \right\}
\end{equation}
where $\psi\left( x\right) $ is the digamma function.

Although now we can express $\ell_{m}$ in the comma basis using the
definitions in (\ref{eqndefofviraopLeft}) and (\ref{eqndefofviraopRight});
it is much easier to obtain $\ell_{m}$ with the aid of the change of
representations relations between the full string and the comma string and
then taking care of any ambiguities that can arise from normal ordering of $%
\alpha^{\prime}s$ and $\beta^{\prime}s$ operators\footnote{%
We have checked that both ways give precisely the same form. We chose to
follow the second path here simply to save space since the second option is
much shorter.} 
\begin{equation*}
\ell_{m}\left( \beta_{q}\right) =\ell_{m}^{\left( L\right) }\left(
\beta_{q}^{\left( \pm\right) }\right) +\ell_{m}^{\left( R\right) }\left(
\beta_{q}^{\left( \pm\right) }\right) \equiv
\end{equation*}%
\begin{equation}
L_{m}\left( \alpha_{q}=\left\{ 
\begin{array}{c}
M_{q\text{ }q^{\prime}}\beta_{q^{\prime}}^{\left( +\right) }+\left(
-1\right) ^{q/2}\wp^{M}\text{ } \\ 
\text{for }q=even\text{, \ }q^{\prime}=\text{odd} \\ 
\alpha_{q}=\beta_{q}^{\left( -\right) }\text{ for }q=\text{odd}%
\end{array}
\right. \right)  \label{eqnconnlhandlf}
\end{equation}
where $\beta^{\left( +\right) }$ has been introduced before and $%
\beta^{\left( -\right) }$ is defined by \ \ 
\begin{equation}
\beta_{2n-1}^{\left( -\right) }=\frac{1}{\sqrt{2}}\left( \beta
_{2n-1}^{\left( L\right) }-\beta_{2n-1}^{\left( R\right) }\right)
\end{equation}
For $\ell_{2k-1}$, skipping tedious but reasonable amount of algebra,
equations (\ref{eqnconnlhandlf}) and (\ref{eq3vgfoeam}) yield 
\begin{equation*}
::\ell_{2k-1}^{\left( L\right) \varphi,i}+\ell_{2k-1}^{\left( R\right)
\varphi,i}::=::\ell_{2k-1}^{\varphi,i}\text{ }::=\lim_{\epsilon\rightarrow
0^{+}}::\ell_{2k-1}^{_{\epsilon}\varphi,i}::
\end{equation*}%
\begin{equation}
=\lim_{\epsilon\rightarrow0}\displaystyle\sum
\limits_{q=even}\beta_{2k-1-q}^{\epsilon\left( -\right) ,i}\left[ %
\displaystyle\sum \limits_{q^{\prime}=odd}M_{q\text{ }q^{\prime}}\beta_{q^{%
\prime}}^{\epsilon\left( +\right) ,i}\right.  \notag
\end{equation}%
\begin{equation}
\left. +\left( -1\right) ^{n/2}\wp_{M}^{\phi,i}\right] -\frac{3}{2}\left(
2k-1\right) \beta_{2k-1}^{\epsilon\left( -\right) ,i}
\end{equation}
where 
\begin{equation}
\beta_{q}^{\left( \pm\right) \epsilon}=e^{-\left\vert \frac{q+1}{2}%
\right\vert \epsilon}\beta_{q}^{\left( \pm\right) }
\end{equation}
The $\ell_{2k-1}^{\varphi}$ for the ghost part differs from the $\ell
_{2k-1}^{\chi}$ for the coordinate part only in the extra terms linear in $%
\beta_{2k-1}^{\epsilon\left( +\right) }$. Simplifying the above expression,
we get%
\begin{equation*}
::\ell_{2k-1}^{\left( L\right) \varphi,i}+\ell_{2k-1}^{\left( R\right)
\varphi,i}::=::\ell_{2k-1}^{\left( \epsilon\right) \varphi,i}::
\end{equation*}%
\begin{equation*}
=\wp_{M}^{\phi.i}\sum_{q\text{ even}}\left( -1\right) ^{q/2}e^{\left\vert 
\frac{2k-q}{2}\right\vert \epsilon}\beta_{2k-1-q}^{\left( -\right) \varphi,i}
\end{equation*}%
\begin{equation*}
+\sum_{q=even\text{ }}\sum_{q^{\prime}=odd\text{ }}G_{k}^{\epsilon}\left(
q,q^{\prime}\right) \beta_{2k-1-q}^{\left( -\right) \varphi,i}\text{ }%
\beta_{q^{\prime}}^{\left( +\right) \varphi,i}\text{ }
\end{equation*}%
\begin{equation}
-\frac{3}{2}\left( 2k-1\right) e^{-\left\vert k\right\vert
\epsilon}\beta_{2k-1}^{\left( -\right) \varphi,i}\text{, \ \ \ }k>0
\label{eqnVirosorGmodd}
\end{equation}
where $G_{k}^{\epsilon}\left( n,q\right) $, a function of the parameter $%
\epsilon$ , is defined by%
\begin{equation}
\text{ }G_{k}^{\epsilon}\left( q,q^{\prime}\right) =e^{-\left\vert \frac{2k-q%
}{2}\right\vert \epsilon}M_{q\text{ }q^{\prime}}e^{-\left\vert \left(
q^{\prime}+1\right) /2\right\vert \epsilon}
\end{equation}
with $q$ even integer and $q^{\prime}$ odd integer. For $m=even=2k>0$, 
\begin{equation}
::\ell_{2k}^{\left( L\right) \varphi,i}+\ell_{2k}^{\left( R\right)
\varphi,i}::=::\ell_{2k}^{\varphi,i}::=\lim_{\epsilon\rightarrow0}::\ell
_{2k}^{\left( \epsilon\right) \varphi,i}::  \label{eqnVirosorXm}
\end{equation}
From the above expression and equations (\ref{eqnconnlhandlf}) and (\ref%
{eq3vgfoeam}), we obtain the even comma Virasoro generators (execluding $m=0$%
) 
\begin{equation*}
::\ell_{2k}^{\left( L\right) \varphi,i}+\ell_{2k}^{\left( R\right)
\varphi,i}::=::\ell_{2k}^{\left( \epsilon\right) \varphi,i}::=\frac{1}{2}%
\sum_{q\text{ odd}}^{\infty}e^{-\left\vert \frac{2k-q-1}{2}\right\vert
\epsilon}
\end{equation*}%
\begin{equation*}
\times e^{-\left\vert \frac{q+1}{2}\right\vert \epsilon}\left[ \beta
_{2k-q}^{\left( -\right) \varphi,i}\beta_{q}^{\left( -\right) \varphi
,i}+\left( -1\right) ^{k}\beta_{2k-q}^{\left( +\right) \varphi,i}\beta
_{q}^{\left( +\right) \varphi,i}\right]
\end{equation*}%
\begin{equation*}
+\left( -1\right) ^{k}\left\{ \left[ \wp_{\phi}^{i}-\frac{3}{2}\left(
2k\right) \right] ^{2}H_{k}^{\epsilon}+2\left[ \wp_{\phi}^{i}-\frac{3}{2}%
\left( 2k\right) \right] \right.
\end{equation*}%
\begin{equation}
\left. \sum_{q\text{ odd}}^{\infty}G_{k}^{\epsilon}\left( q\right)
\beta_{q}^{\left( +\right) \varphi,i}+\sum_{q,q^{\prime}\text{ odd}%
}^{\infty}F_{k}^{\epsilon}\left( q,q^{\prime}\right) \beta_{q}^{\left(
+\right) \varphi,i}\beta_{q^{\prime}}^{\left( +\right) \varphi,i}\text{ }%
\right\}
\end{equation}
where the quantities, $H$, $G$, and $F$, are given by 
\begin{equation}
\text{ }H_{k}^{\epsilon}=\sum_{q^{\prime\prime}\text{ even}}^{\infty
}e^{-\left\vert \frac{2k-q^{\prime\prime}}{2}\right\vert \epsilon
}e^{-\left\vert \frac{q^{\prime\prime}}{2}\right\vert \epsilon}
\end{equation}%
\begin{equation}
G_{k}^{\epsilon}\left( q\right) =\sum_{q^{\prime\prime}\text{ even}%
}^{\infty}e^{-\left\vert \frac{q}{2}\right\vert \epsilon}e^{-\left\vert 
\frac{q^{\prime\prime}}{2}\right\vert \epsilon}G_{k}^{\epsilon}\left(
q,q^{\prime\prime}\right)
\end{equation}%
\begin{equation*}
F_{k}^{\epsilon}\left( q,q^{\prime}\right) =\left( -1\right) ^{k}\frac {2}{%
\pi}\frac{\left( -1\right) ^{\frac{q+1}{2}}}{q}G_{k}^{\epsilon}\left(
q^{\prime}\right)
\end{equation*}%
\begin{equation*}
+\frac{1}{2}\sum_{q^{\prime\prime}\text{ even}}^{\infty}\left[
1-e^{\left\vert \frac{q^{\prime\prime}}{2}\right\vert
\epsilon}e^{-\left\vert \frac{2k-q^{\prime\prime}}{2}\right\vert \epsilon}%
\right]
\end{equation*}%
\begin{equation}
\times\frac{\left( -1\right) ^{k}}{2k-q^{\prime\prime}}G_{k}^{\epsilon
}\left( q^{\prime\prime},q\right) G_{k}^{\epsilon}\left( 2k-q^{\prime
\prime},q^{\prime}\right)
\end{equation}
The comma Virasoro generators obtain by taking the limit in $\epsilon$. To
first order in $\epsilon$ one obtains%
\begin{equation*}
::\ell_{2k}^{\left( L\right) \varphi,i}+\ell_{2k}^{\left( R\right)
\varphi,i}::=::\ell_{2k}^{\varphi,i}::=\lim_{\epsilon\rightarrow0}::\ell
_{2k}^{\left( \epsilon\right) \varphi,i}::
\end{equation*}%
\begin{equation*}
=\frac{1}{2}\sum_{q\text{ odd}}^{\infty}\left[ \beta_{2k-q}^{\left( -\right)
\varphi,i}\beta_{q}^{\left( -\right) \varphi,i}+\left( -1\right)
^{k}\beta_{2k-q}^{\left( +\right) \varphi,i}\beta_{q}^{\left( +\right)
\varphi,i}\right] +
\end{equation*}%
\begin{equation*}
\lim_{\epsilon\rightarrow0}\frac{\left( -1\right) ^{k}}{2\epsilon}\left\{ %
\left[ \wp^{i}-\frac{3}{2}\left( 2k\right) \right] ^{2}+2\left[ \wp^{i}-%
\frac{3}{2}\left( 2k\right) \right] \frac{2}{\pi}\right. \sum_{q\text{ odd}%
}^{\infty}
\end{equation*}%
\begin{equation*}
\frac{\left( -1\right) ^{\frac{q+1}{2}}}{q}e^{-\left\vert \frac{q+1}{2}%
\right\vert \epsilon}\beta_{q}^{\left( +\right) \varphi,i}+\left( \frac{2}{%
\pi}\right) ^{2}\sum_{q,q^{\prime}\text{ odd}}^{\infty}\frac{\left(
-1\right) ^{\frac{q+1}{2}}}{q}
\end{equation*}%
\begin{equation}
\text{ }\left. \frac{\left( -1\right) ^{\frac{q^{\prime}+1}{2}}}{q^{\prime }}%
e^{-\left\vert \frac{q+1}{2}\right\vert \epsilon}e^{-\left\vert \frac{%
q^{\prime}+1}{2}\right\vert \epsilon}\beta_{q}^{\left( +\right)
\varphi,i}\beta_{q^{\prime}}^{\left( +\right) \varphi,i}\right\}
\label{eqnVirosorGmeven}
\end{equation}
Likewise for $\ell_{0}^{\varphi,i}$ one obtains \ \ \ 

\begin{equation*}
\text{ }::\ell_{0}^{\left( L\right) \varphi,i}+\ell_{0}^{\left( R\right)
\varphi,i}::=::\ell_{0}^{\varphi,i}::
\end{equation*}%
\begin{equation*}
=\frac{1}{2}\sum_{q\text{ odd}}^{\infty}\left[ \beta_{-q}^{\left( -\right)
\varphi,i}\beta_{q}^{\left( -\right) \varphi,i}+\beta_{-q}^{\left( +\right)
\varphi,i}\beta_{q}^{\left( +\right) \varphi,i}\right] -\frac {1}{8}
\end{equation*}%
\begin{align*}
\text{ } & +\lim_{\epsilon\rightarrow0}\frac{1}{2\epsilon}\left\{ \left(
\wp_{\phi}^{i}\right) ^{2}-1+2\wp_{\phi}^{i}\frac{2}{\pi}\sum_{q\text{ odd}%
}^{\infty}\frac{\left( -1\right) ^{\frac{q+1}{2}}}{q}\right. \\
& \times e^{-\left\vert \frac{q+1}{2}\right\vert \epsilon}\beta_{q}^{\left(
+\right) \varphi,i}+\left( \frac{2}{\pi}\right) ^{2}\sum_{q,q^{\prime }\text{
odd}}^{\infty}\frac{\left( -1\right) ^{\frac{q+1}{2}}}{q}
\end{align*}%
\begin{equation}
\left. \times\frac{\left( -1\right) ^{\frac{q^{\prime}+1}{2}}}{q^{\prime}}%
e^{-\left\vert \frac{q+1}{2}\right\vert \epsilon}e^{-\left\vert \frac{%
q^{\prime}+1}{2}\right\vert \epsilon}\beta_{q}^{\left( +\right)
\varphi,i}\beta_{q^{\prime}}^{\left( +\right) \varphi,i}\right\}
\label{eqnVirosorGmzero}
\end{equation}
The extra term in (\ref{eqnVirosorGmeven}), linear in $\beta^{\left(
+\right) }$, arises because of the presence of the $R\phi$ term in the
action of the bosonized ghosts \cite{E.Witten-CST} 
\begin{equation}
I_{\phi}=\frac{1}{2\pi}\int d^{2}\sigma\left( \partial_{\lambda}\phi
\partial^{\lambda}\phi-3iR\phi\right)
\end{equation}
or alternatively because of the extra linear term in the ghost
energy-momentum tensor (\ref{eqenergymomtensorbose}). The extra term is
needed and must have precisely the coefficient given in (\ref%
{eqenergymomtensorbose}), so that $\phi$ (ghost coordinate) can cancel the
Virasoro anomaly of the $x^{\mu}$ (matter coordinates) so that the total
(matter plus ghost) Fourier components of the energy momentum%
\begin{equation}
\ell_{m}^{\left( \epsilon\right) s}=\ell_{m}^{\left( \epsilon\right)
\chi,s}+\ell_{m}^{\left( \epsilon\right) \varphi,s}-\frac{9}{8}\delta_{m0}
\end{equation}
obey the Virasoro algebra%
\begin{equation}
\left[ \ell_{m}^{\epsilon,s},\ell_{n}^{\epsilon,s}\right] =\left( m-n\right)
\ell_{m+n}^{\epsilon,r}
\end{equation}
which is free of central charge.

Utilizing the relationship between the two normal orderings, we can easily
deduce the relationship between the Virasoro generators in both cases 
\begin{equation}
::\ell_{2k-1}^{\left( \epsilon\right) \chi,\varphi}::=:\ell_{2k-1}^{\left(
\epsilon\right) \chi+\varphi}:-\frac{1}{4\epsilon}\left( \frac{1}{\epsilon }%
+\log\frac{\epsilon}{2}+1\right)
\end{equation}
where we have kept only divergent terms of order $\epsilon$.

Now we are in position to write down the explicit form of the $\mathcal{WL}$
operator in the comma basis. We recall that%
\begin{equation*}
::\mathcal{W}_{m}^{\chi+\varphi,i}\left( L\right) +\mathcal{W}%
_{m}^{\chi+\varphi,i}\left( R\right) ::=::W_{m}^{\chi+\varphi,i}::
\end{equation*}%
\begin{equation*}
=::\ell_{m}^{\left( L\right) \chi+\varphi,i}+\ell_{m}^{\left( R\right)
\chi+\varphi,i}::+\sum_{j=1}^{3}
\end{equation*}%
\begin{equation}
\sum_{n=0}^{\infty}m\widetilde{N}_{mn}^{ij}::\ell_{-n}^{\left( L\right)
\chi+\varphi,j}+\ell_{-n}^{\left( R\right) \chi+\varphi,j}::
\label{eqndefwaropincom}
\end{equation}
with $\ell_{-n}\equiv\ell_{n}^{\dag}$.

Let us here focus on the ghost part since the coordinate part will have the
same structure with the absence of terms linear \ in the comma creation and
annihilation operators. For $m=odd=2k-1>0$, equation (\ref{eqndefwaropincom}%
) gives%
\begin{equation*}
::\mathcal{W}_{2k-1}^{\varphi,i}\left( L\right) +\mathcal{W}%
_{2k-1}^{\varphi,i}\left( R\right) ::=::W_{2k-1}^{\varphi,i}::
\end{equation*}%
\begin{equation*}
::=::\ell_{2k-1}^{\left( L\right) \varphi,i}+\ell_{2k-1}^{\left( R\right)
\varphi,i}::+\sum_{j=1}^{3}\sum_{n=0}^{\infty}\left( 2k-1\right)
\end{equation*}%
\begin{equation}
\times\widetilde{N}_{2k-1\text{ }n}^{ij}::\ell_{-n}^{\left( L\right)
\varphi,j}+\ell_{-n}^{\left( R\right) \varphi,j}::
\end{equation}
Substituting (\ref{eqnVirosorGmodd}), (\ref{eqnVirosorGmeven}) and (\ref%
{eqnVirosorGmzero}) into the above expression, after a lengthy algebra, we
obtain%
\begin{equation*}
::\mathcal{W}_{2k-1}^{\varphi,i}\left( L\right) +W_{2k-1}^{\varphi,i}\left(
R\right) ::=::W_{2k-1}^{\varphi,i}::
\end{equation*}%
\begin{equation*}
=\wp_{M}^{\phi,j}\sum_{q\text{ even}}\left( -1\right) ^{\frac{q}{2}%
}e^{\left\vert \frac{2k-q}{2}\right\vert \epsilon}\beta_{2k-1-q}^{\left(
-\right) \varphi,j}
\end{equation*}%
\begin{equation*}
+\sum_{q=even\text{ }}\sum_{q^{\prime}=odd\text{ }}G_{k}^{\epsilon}\left(
q,q^{\prime}\right) \beta_{2k-1-q}^{\left( -\right) \varphi,j}\text{ }%
\beta_{q^{\prime}}^{\left( +\right) \varphi,j}
\end{equation*}%
\begin{equation*}
-\frac{3}{2}\left( 2k-1\right) e^{-\left\vert k\right\vert
\epsilon}\beta_{2k-1}^{\left( -\right) \varphi,j}+\sum_{j=1}^{3}\left(
2k-1\right) \widetilde{N}_{2k-1\text{ }0}^{ij}
\end{equation*}%
\begin{equation*}
\times\left\{ \frac{1}{2}\sum_{q\text{ odd}}^{\infty}\left[ \beta
_{q}^{\left( -\right) \varphi,j}\beta_{-q}^{\left( -\right) \varphi
,j}+\beta_{q}^{\left( +\right) \varphi,j}\beta_{-q}^{\left( +\right)
\varphi,j}\right] \right. -\frac{1}{8}
\end{equation*}%
\begin{equation*}
+\lim_{\epsilon\rightarrow0}\frac{1}{2\epsilon}\left[ \left(
\wp_{\phi}^{j}\right) ^{2}-1+2\wp_{\phi}^{j}\frac{2}{\pi}\sum_{q\text{ odd}%
}^{\infty}\frac{\left( -1\right) ^{\frac{q+1}{2}}}{q}\right.
\end{equation*}%
\begin{equation*}
\times e^{-\left\vert \frac{q+1}{2}\right\vert \epsilon}\beta_{-q}^{\left(
+\right) \varphi,j}+\left( \frac{2}{\pi}\right) ^{2}\sum_{q,q^{\prime }\text{
odd}}^{\infty}\frac{\left( -1\right) ^{\frac{q+1}{2}}}{q}
\end{equation*}%
\begin{equation*}
\left\vert \left. \times\frac{\left( -1\right) ^{\frac{q^{\prime}+1}{2}}}{%
q^{\prime}}e^{-\left\vert \frac{q+1}{2}\right\vert \epsilon}e^{-\left\vert
\left( q^{\prime}+1\right) /2\right\vert
\epsilon}\beta_{-q^{\prime}}^{\left( +\right) \varphi,j}\beta_{-q}^{\left(
+\right) \varphi ,j}\right] \right\}
\end{equation*}%
\begin{equation*}
+\sum_{j=1}^{3}\sum_{n=1}^{\infty}\left( 2k-1\right) \widetilde{N}_{2k-1%
\text{ }2n}^{ij}\left\{ \frac{1}{2}\sum_{q\text{ odd}}^{\infty
}e^{-\left\vert \frac{2n-q-1}{2}\right\vert \epsilon}\times\right.
\end{equation*}%
\begin{equation*}
e^{-\left\vert \frac{q+1}{2}\right\vert \epsilon}\left[ \beta_{-q}^{\left(
-\right) \varphi,j}\beta_{-\left( 2n-q\right) }^{\left( -\right)
\varphi,j}+\left( -1\right) ^{n}\beta_{-q}^{\left( +\right) \varphi
,j}\beta_{-\left( 2n-q\right) }^{\left( +\right) \varphi,j}\right]
\end{equation*}%
\begin{equation*}
+\left( -1\right) ^{k}\left[ \left( \wp_{\phi}^{j}-\frac{3}{2}\left(
2n\right) \right) ^{2}H_{n}^{\epsilon}+2\times\right.
\end{equation*}

\begin{equation*}
\left( \wp_{\phi}^{j}-\frac{3}{2}\left( 2n\right) \right) \sum_{q\text{ odd}%
}^{\infty}G_{n}^{\epsilon}\left( q\right) \beta _{-q}^{\left( +\right)
\varphi,j}
\end{equation*}

\begin{equation*}
\left. \left. +\sum_{q,q^{\prime}\text{ odd}}^{\infty}F_{n}^{\epsilon
}\left( q,q^{\prime}\right) \beta_{-q^{\prime}}^{\left( +\right) \varphi,j}%
\text{ }\beta_{-q}^{\left( +\right) \varphi,j}\right] \right\}
\end{equation*}%
\begin{equation*}
+\sum_{j=1}^{3}\sum_{n=1}^{\infty}\left( 2k-1\right) \widetilde{N}_{2k-1%
\text{ }2n-1}^{ij}\left\{ \wp_{M}^{\phi,j}\sum_{q\text{ even}}\left(
-1\right) \right.
\end{equation*}%
\begin{equation*}
^{q/2}e^{\left\vert \left( 2n-q\right) /2\right\vert \epsilon}\beta_{-\left(
2n-1-q\right) }^{\left( -\right) \varphi,j}+\sum _{q=even\text{ }%
}\sum_{q^{\prime}=odd\text{ }}G_{n}^{\epsilon}\left( q,q^{\prime}\right)
\end{equation*}

\begin{equation}
\left. \beta_{-q^{\prime}}^{\left( +\right) \varphi,j}\text{ }\beta_{-\left(
2n-1-q\right) }^{\left( -\right) \varphi,j}-\frac{3}{2}\left( 2n-1\right)
e^{-\left\vert n\right\vert \epsilon}\beta_{-\left( 2n-1\right) }^{\left(
-\right) \varphi,j}\right\}
\end{equation}

\bigskip Likewise for $m=even=2k>0$, equation (\ref{eqndefwaropincom}) yields%
\begin{equation*}
::\mathcal{W}_{2k}^{\varphi,i}\left( L\right) +\mathcal{W}_{2k}^{\varphi
,i}\left( R\right) ::=::W_{2k}^{\varphi,i}::=::\ell_{2k}^{\left( L\right)
\varphi,i}
\end{equation*}%
\begin{equation}
+\ell_{2k}^{\left( R\right) \varphi,i}::+\sum_{j=1}^{3}\sum_{n=0}^{\infty }2k%
\widetilde{N}_{2k\text{ }n}^{ij}::\ell_{-n}^{\left( L\right) \varphi
,j}+\ell_{-n}^{\left( R\right) \varphi,j}::
\end{equation}
Substituting equations (\ref{eqnVirosorGmodd}), (\ref{eqnVirosorGmeven}) and
(\ref{eqnVirosorGmzero}) into the above expression, after a lengthy algebra,
we get%
\begin{equation*}
::\mathcal{W}_{2k}^{\varphi,i}\left( L\right) +\mathcal{W}_{2k}^{\varphi
,i}\left( R\right) ::=::W_{2k}^{\varphi,i}::
\end{equation*}%
\begin{equation*}
=\frac{1}{2}\sum_{q\text{ odd}}^{\infty}\left[ \beta_{2k-q}^{\left( -\right)
\varphi,i}\beta_{q}^{\left( -\right) \varphi,i}+\left( -1\right)
^{k}\beta_{2k-q}^{\left( +\right) \varphi,i}\beta_{q}^{\left( +\right)
\varphi,i}\right]
\end{equation*}%
\begin{equation*}
+\left( -1\right) ^{k}\lim_{\epsilon\rightarrow0}\frac{1}{2\epsilon}\left( %
\left[ \wp^{i}-\frac{3}{2}\left( 2k\right) \right] ^{2}+2\left[ \wp^{i}-%
\frac{3}{2}\left( 2k\right) \right] \right.
\end{equation*}%
\begin{equation*}
\times\frac{2}{\pi}\sum_{q\text{ odd}}^{\infty}\frac{\left( -1\right) ^{%
\frac{q+1}{2}}}{q}e^{-\left\vert \frac{q+1}{2}\right\vert \epsilon}\beta
_{q}^{\left( +\right) \varphi,i}+\left( \frac{2}{\pi}\right)
^{2}\sum_{q,q^{\prime}\text{ odd}}^{\infty}
\end{equation*}%
\begin{equation*}
\left. \frac{\left( -1\right) ^{\frac{q+1}{2}}}{q}\frac{\left( -1\right) ^{%
\frac{q^{\prime}+1}{2}}}{q^{\prime}}e^{-\left\vert \frac{q+1}{2}\right\vert
\epsilon}e^{-\left\vert \frac{q^{\prime}+1}{2}\right\vert \epsilon}\beta
_{q}^{\left( +\right) \varphi,i}\beta_{q^{\prime}}^{\left( +\right)
\varphi,i}\right)
\end{equation*}%
\begin{equation*}
+\sum_{j=1}^{3}2k\widetilde{N}_{2k0}^{ij}\left\{ \frac{1}{2}\sum_{q\text{ odd%
}}^{\infty}\left[ \beta_{-q}^{\left( -\right) \varphi,j}\beta _{q}^{\left(
-\right) \varphi,j}\right. \right.
\end{equation*}%
\begin{equation*}
\left. +\beta_{-q}^{\left( +\right) \varphi,j}\beta_{q}^{\left( +\right)
\varphi,j}\right] -\frac{1}{8}+\lim_{\epsilon\rightarrow0}\frac{1}{2\epsilon}
\end{equation*}%
\begin{equation*}
\left( \left( \wp_{\phi}^{j}\right) ^{2}-1+2\wp_{\phi}^{j}\frac{2}{\pi}%
\sum_{q\text{ odd}}^{\infty}\frac{\left( -1\right) ^{\frac{q+1}{2}}}{q}%
\right.
\end{equation*}%
\begin{equation*}
\times e^{-\left\vert \frac{q+1}{2}\right\vert \epsilon}\beta_{-q}^{\left(
+\right) \varphi,j}+\left( \frac{2}{\pi}\right) ^{2}\sum_{q,q^{\prime }\text{
odd}}^{\infty}\frac{\left( -1\right) ^{\frac{q+1}{2}}}{q}\times
\end{equation*}%
\begin{equation*}
\left. \left. \frac{\left( -1\right) ^{\frac{q^{\prime}+1}{2}}}{q^{\prime }}%
e^{-\left\vert \left( \frac{q+1}{2}\right) \right\vert \epsilon
}e^{-\left\vert \frac{q^{\prime}+1}{2}\right\vert
\epsilon}\beta_{-q^{\prime}}^{\left( +\right) \varphi,j}\beta_{-q}^{\left(
+\right) \varphi ,j}\right) \right\}
\end{equation*}%
\begin{equation*}
+\sum_{j=1}^{3}\sum_{n=1}^{\infty}2k\widetilde{N}_{2k\text{ }2n}^{ij}\left\{ 
\frac{1}{2}\sum_{q\text{ odd}}^{\infty}e^{-\left\vert \frac{2n-q-1}{2}%
\right\vert \epsilon}e^{-\left\vert \frac{q+1}{2}\right\vert \epsilon
}\right.
\end{equation*}%
\begin{equation*}
\times\left[ \beta_{-q}^{\left( -\right) \varphi,j}\beta_{-\left(
2n-q\right) }^{\left( -\right) \varphi,j}+\left( -1\right) ^{n}\beta
_{-q}^{\left( +\right) \varphi,j}\beta_{-\left( 2n-q\right) }^{\left(
+\right) \varphi,j}\right]
\end{equation*}%
\begin{equation*}
+\left( -1\right) ^{k}\left[ \left( \wp_{\phi}^{j}-\frac{3}{2}\left(
2n\right) \right) ^{2}H_{n}^{\epsilon}+2\left( \wp_{\phi}^{j}-\frac{3}{2}%
\left( 2n\right) \right) \times\right.
\end{equation*}%
\begin{equation*}
\sum_{q\text{ odd}}^{\infty}G_{n}^{\epsilon}\left( q\right) \beta
_{-q}^{\left( +\right) \varphi,j}\left. \left. +\sum_{q,q^{\prime}\text{ odd}%
}^{\infty}F_{n}^{\epsilon}\left( q,q^{\prime}\right)
\beta_{-q^{\prime}}^{\left( +\right) \varphi,j}\text{ }\beta_{-q}^{\left(
+\right) \varphi,j}\right] \right\}
\end{equation*}%
\begin{equation*}
+\sum_{j=1}^{3}\sum_{n=1}^{\infty}2k\widetilde{N}_{2k\text{ }\left(
2n-1\right) }^{ij}\left\{ \wp_{M}^{\phi,j}\sum_{q\text{ even}}\left(
-1\right) ^{\frac{q}{2}}e^{\left\vert \frac{2n-q}{2}\right\vert \epsilon
}\right.
\end{equation*}%
\begin{equation*}
\times\beta_{-\left( 2n-1-q\right) }^{\left( -\right) \varphi,j}+\sum_{q=even%
\text{ }}\sum_{q^{\prime}=odd\text{ }}G_{n}^{\epsilon}\left(
q,q^{\prime}\right) \beta_{-q^{\prime}}^{\left( +\right) \varphi,j}\text{ }
\end{equation*}%
\begin{equation}
\beta_{-\left( 2n-1-q\right) }^{\left( -\right) \varphi,j}\left. -\frac{3}{2}%
\left( 2n-1\right) e^{-\left\vert n\right\vert \epsilon}\beta_{-\left(
2n-1\right) }^{\left( -\right) \varphi,j}\right\}
\end{equation}
And last for $m=0$, equation (\ref{eqndefwaropincom}) yields%
\begin{equation*}
\text{ }::\mathcal{W}_{0}^{\varphi,i}\left( L\right) +\mathcal{W}%
_{0}^{\varphi,i}\left( R\right) ::
\end{equation*}%
\begin{equation*}
\text{ }=::\mathcal{W}_{0}^{\chi+\varphi,i}:=::\ell_{0}^{\left( L\right)
\varphi,i}+\ell_{0}^{\left( R\right) \varphi,i}::
\end{equation*}%
\begin{equation*}
\text{ }=\frac{1}{2}\sum_{q\text{ odd}}^{\infty}\left[ \beta_{-q}^{\left(
-\right) \varphi,i}\beta_{q}^{\left( -\right) \varphi,i}+\beta _{-q}^{\left(
+\right) \varphi,i}\beta_{q}^{\left( +\right) \varphi ,i}\right] -\frac{1}{8}
\end{equation*}%
\begin{equation*}
+\lim_{\epsilon\rightarrow0}\frac{1}{2\epsilon}\left\{ \left(
\wp_{\phi}^{i}\right) ^{2}-1+2\wp_{\phi}^{i}\frac{2}{\pi}\sum_{q\text{ odd}%
}^{\infty}\frac{\left( -1\right) ^{\frac{q+1}{2}}}{q}\right.
\end{equation*}%
\begin{equation*}
\times e^{-\left\vert \frac{q+1}{2}\right\vert \epsilon}\beta_{q}^{\left(
+\right) \varphi,i}+\left( \frac{2}{\pi}\right) ^{2}\sum_{q,q^{\prime }\text{
odd}}^{\infty}\frac{\left( -1\right) ^{\frac{q+1}{2}}}{q}\times
\end{equation*}%
\begin{equation}
\left. \frac{\left( -1\right) ^{\frac{q^{\prime}+1}{2}}}{q^{\prime}}%
e^{-\left\vert \frac{q+1}{2}\right\vert \epsilon}e^{-\left\vert \frac{%
q^{\prime}+1}{2}\right\vert \epsilon}\beta_{q}^{\left( +\right)
\varphi,i}\beta_{q^{\prime}}^{\left( +\right) \varphi,i}\right\}
\end{equation}
The corresponding coordinates parts of the $\mathcal{WL}$ operator are given
by the same expressions with the terms linear in $\beta^{\prime}s$ deleted.

\section{Conclusions}

We have seen in \cite{Bordes} that the comma vertex can be expressed in
terms of bose or fermi fields. Part of the proof of equivalence of the two
forms requires finding the $\mathcal{WL}$ operator in the comma theory. We
have constructed the $\mathcal{WL}$ operator in the comma formulation of
Witten's open bosonic string which was the main obstacle in proving the $%
\mathcal{WL}$ identities. The proof of the $\mathcal{WL}$ identities and $%
BRST$- invariance are given in a sequel to this paper \cite{AbduWardId}.

\end{document}